\documentstyle[12pt]{article}
\normalbaselines
\begin{document}

\baselineskip=24pt

\bibliographystyle{unsrt}
\vbox {\vspace{6mm}}

\begin{center} {\bf RECONSTRUCTING THE DENSITY OPERATOR\\ BY USING GENERALIZED FIELD QUADRATURES}
\end{center}        

\bigskip

\begin{center} 

{\it G. M. D'Ariano }$^{\dag}$\footnote{Also at Istituto Nazionale di Fisica della Materia and
Istituto Nazionale di Fisica Nucleare, Italy.}, {\it S. Mancini }$^{* 1}$,  {\it V. I.
Man'ko}$^{\ddag}$   {\it and P. Tombesi }$^{* 1}$

\end{center}

\bigskip

\begin{center}

$^{\dag}$ Dipartimento di Fisica "A. Volta", Universit\`a di Pavia, I-27100 Pavia, Italy\\
$^*$ Dipartimento di Matematica e Fisica, Universit\`a di Camerino, I-62032 Camerino, Italy\\
$^{\ddag}$ Lebedev Physical Institute, Leninsky Prospect 53, 117924 Moscow Russia

\end{center}

\bigskip

\bigskip

\begin{abstract}

\baselineskip=24pt

The Wigner function for one and two-mode quantum systems is explicitely expressed in terms of the
marginal distribution for the generic linearly transformed quadratures. Then, also the density
operator of those systems is written in terms of the marginal distribution of these quadratures. Some
examples to apply this formalism, and a reduction to the usual optical homodyne tomography are
considered.

\end{abstract}

PACS number(s): 42.50.Dv, 03.65.Bz

\bigskip
\bigskip
\bigskip
\bigskip

\section{Introduction}\label{s1} 

The homodyne measurement of an electromagnetic field gives all the possible 
combinations of the field quadratures belonging to the group $O(2)$, by just varying the phase of
the local oscillator. The average of the random outcomes of the measurement, at a given local
oscillator phase, is connected with the marginal distribution of the Wigner function, or any other
quasi-probability used in quantum optics. In the work \cite{vogel} it was shown, indeed, that the
rotated quadrature distribution may be expressed in terms of the Wigner function \cite{wigner32} as
well as in terms of the Husimi
$Q$-function
\cite{hus40} and the Glauber-Sudarshan $P$-representation \cite{gla63}, \cite{sud63}. The result of
Ref.
\cite{vogel} was based on the observation that relations of the density matrix in different
representations to the characteristic function, intended as the mean value of the displacement
operator creating a coherent state from the vacuum \cite{gla63}, \cite{cahgla69},
may be rewritten as relations where the characteristic function is the mean value of the
rotated quadrature phase. It gave the possibility to express the Wigner function in terms of the
marginal distribution of homodyne outcomes through the tomographic formula. The essential point of
the obtained formula is that the homodyne output distribution may be measured and the corresponding
inverse Radon transform produces the Wigner distribution function, which is refered as the optical
homodyne tomography \cite{raimer}. The density matrix elements, in some representations, can also be
obtained by avoiding the Wigner function and then the Radon transform \cite{wvogel},
\cite{darianopar},
\cite{APL}.

The tomographic formula of Ref. [1] has been presented in an invariant form in Ref. \cite{APL}. In
that work the system density operator was expressed as convolution of the marginal distribution of
the homodyne output and a kernel operator. In Ref. \cite{MMT} the tomographic formula of Ref. [1]
was extended to express the Wigner function in terms of the marginal distribution of the
generalized quadrature obtained through canonical transforms belonging to the Lie group
$ISp(2,R)$. Essentially this generalized quadrature differs from the one used in the standard
tomography since it depends upon three real parameters instead of only one (really, a scaling
transformation is also present in the homodyne measurement because position and momentum have
different units, but the scaling cannot be independently controlled  of the rotation). 

The aim of this paper is to obtain the invariant form of the
density operator of the quantum system in terms of the discussed \cite{MMT} marginal distributions
for the generalized quadratures, which is analogous to Ref. \cite{APL}, and to extend the
analysis to the multimode case. 

Recently the quantum tomography was also considered to investigate
material systems
\cite{wal} besides the optical ones, thus our alternative approach could be useful for a better
characterization of their quantum state since it allows to get overcomplete information. 

The structure of the article is the following. In Section 2 we derive the density operator as a
convolution of the marginal distribution for the generic linear quadrature and a kernel operator,
and discuss the properties of the obtained formula in number state basis, coherent states basis and
coordinate representation. In Section 3 we consider the obtained formula applied to some
examples of interest, such as the ground state of a harmonic oscillator, the
Schr\"odinger-cat state, and the thermal equilibrium state of an
oscillator. In Section 4 we contrast the obtained formula with the tomographic one for the density
operator of Ref.
\cite{APL}. In Sections 5 and 6 the approach will be  extended to the multimode case. 
Some applications of the developped formalism as well as
practical implementation and further developments will be discussed in Sections 7 and 8.

\section{Density Matrix and Marginal Distribution}

Let us consider the quadrature observable $\hat X$, which is a generic linear form in position $\hat
q$ and momentum $\hat p$
\begin{equation}\label{quadef}
\hat X=\mu \hat q+\nu \hat p+\delta
\end{equation}
with $\mu,\nu,\delta$ real parameters (their physical meaning will be discuss later on), then
it is possible to get the density matrix elements from the marginal distribution avoiding the
evaluation of the Wigner function as an intermediate step. For this pourpose we start from a well
known \cite{cahgla69} representation of the density operator  
\begin{equation}\label{rho1}
\hat\rho=\int \frac{d^2\alpha}{\pi} W(\alpha)\hat T(\alpha)
\end{equation}
where the Wigner function $W(\alpha)$ is a weight function for the expansion of the density
operator in terms of the operator $\hat T(\alpha)$ which is defined as the complex Fourier
transform of the displacement operator $\hat D$
\begin{equation}\label{Tdef}
\hat T(\alpha)=\int \frac{d^2\xi}{\pi}\hat D(\xi)\exp(\alpha\xi^*-\alpha^*\xi).
\end{equation}
Following the lines of Ref. \cite{MMT}
we may write the marginal distribution $w$ for the generic
quadrature of Eq. (\ref{quadef}) as
\begin{equation}\label{margXdelta}
w(X,\mu,\nu,\delta)=\int e^{-ik(X-\mu q-\nu p-\delta)}
W(q,p)\frac{dkdqdp}{(2\pi)^2}\,,
\end{equation}
where the canonical coordinate $q$ and $p$ are related to $\alpha$ by 
\begin{equation}\label{alphadef}
\alpha=\frac{q+ip}{\sqrt{2}}\,.
\end{equation}
Eq. (\ref{margXdelta}) shows that $w$ is function of the difference $X-\delta=x$, so that it can be
rewritten as
\begin{equation}\label{margx}
w(x,\mu,\nu)=\int e^{-ik(x-\mu q-\nu p)}
W(q,p)\frac{dkdqdp}{(2\pi)^2}\,;
\end{equation}
and by means of the Fourier transform of the function $w$ one can obtain the relation
\begin{equation}\label{wig}
W(q,p)=(2\pi)^2z^2w(z,-zq,-zp)
\end{equation}
where $-zq,\,-zp,\,z$ are the conjugate variable to $\mu,\,\nu,\,x$
respectively. 
It is worth to remark that in this case the connection between the Wigner function and the marginal
distribution is simply guaranteed by means of Fourier transform instead of the Radon one. Then
inserting Eq. (\ref{wig}) into Eq. (\ref{rho1}), expressing the marginal distribution in terms of the
Fourier transform, we have:
\begin{equation}\label{denop}
\hat\rho=\int\frac{dqdp}{(2\pi)^2}\hat T(q,p)\int dx d\mu d\nu \quad
z^2w(x,\mu,\nu)e^{-izx+i\mu zq+i\nu zp},
\end{equation}
or, in a compact form 
\begin{equation}\label{density}
\hat\rho=\int dx d\mu d\nu \quad w(x,\mu,\nu)\hat K_{\mu,\nu}\,,
\end{equation}
where the kernel operator $\hat K_{\mu,\nu}$ is given by:
\begin{eqnarray}\label{ker}
\hat K_{\mu,\nu}&=&\int\frac{dqdp}{(2\pi)^2}z^2e^{+i\mu zq+i\nu zp-izx}\hat T(q,p)
\nonumber\\
&=&\frac{1}{2\pi}z^2e^{-izx}e^{-\frac{z}{\sqrt{2}}(\nu-i\mu)\hat a^{\dag}}
e^{\frac{z}{\sqrt{2}}(\nu+i\mu)\hat a}e^{-\frac{z^2}{4}(\mu^2+\nu^2)}
\end{eqnarray}
where we have used Eq. (\ref{Tdef}) and the Baker-Hausdorff formula.
The fact that $\hat K_{\mu,\nu}$ depends on the $z$ variable as well (i.e. each Fourier component
gives a selfconsistent kernel) shows the overcompleteness of information achievable by measuring the
observable of Eq. (\ref{quadef}). Thus, Eq. (\ref{density}) can be useful to completely determine the
properties of the system, i. e. the density operator, from the probability distribution of the
experimental data; outcomes of a set of measurements such as described in Ref. \cite{MMT}.

Let us now consider the expression of the kernel operator (\ref{ker}) in various
representations. First, in the coherent states basis we have
\begin{equation}\label{kercoh}
\langle\alpha|\hat K_{\mu,\nu}|\beta\rangle=
\frac{1}{2\pi}z^2e^{-izx}e^{-\frac{z}{\sqrt{2}}(\nu-i\mu)\alpha^*}
e^{\frac{z}{\sqrt{2}}(\nu+i\mu)\beta}e^{-\frac{z^2}{4}(\mu^2+\nu^2)}
e^{-\frac{|\alpha|^2}{2}-\frac{|\beta|^2}{2}+\alpha^*\beta},
\end{equation}
while in the number states representation we get
\begin{eqnarray}\label{kernum}
\langle n+d|\hat K_{\mu,\nu}|n\rangle&=&
\frac{1}{2\pi}z^2e^{-izx}e^{-\frac{z^2}{4}(\mu^2+\nu^2)}\nonumber\\
&\times&
\sum_{l=0}^{n}\frac{\sqrt{n!(n+d)!}}{(n-l)!}\frac{[-\frac{z}
{\sqrt{2}}(\nu-i\mu)]^l}{l!}
\frac{[\frac{z}{\sqrt{2}}(\nu+i\mu)]^{(l+d)}}{(l+d)!},
\end{eqnarray}
and finally in the coordinate representation we obtain
\begin{equation}\label{kercoo}
\langle q'|\hat K_{\mu,\nu}|q''\rangle=
\frac{1}{2\pi}z^2e^{-izx}e^{iz^2\mu\nu/2}e^{iz\mu q'}\delta (q''-z\nu-q').
\end{equation}
If, and only if, the kernel operator is bounded every moment of the kernel is bounded for
all possible distributions $w(x,\mu,\nu)$, then, according to the central limit theorem, the
matrix elements of Eq. (\ref{density}) can be sampled on a sufficiently large sets of data only in
the number or coherent states basis, as can be evicted from the expressions (\ref{kercoh}),
(\ref{kernum}) and (\ref{kercoo}). That is in perfect agreement with the results of
Ref. \cite{APL}. 

\section{Examples}

In this Section we will consider some examples of interest to check the validity of Eq.
(\ref{density}). We first consider the case of the harmonic oscillator's ground state for
which we have
\cite{MMT}:
\begin{equation}\label{margs} 
w(x,\mu ,\nu)=[\pi (\mu +\nu )]^{-1/2}\exp [-\frac {
x^2}{\mu ^{2}+\nu ^{2}}].
\end{equation}
In the coherent states basis of Eq. (\ref{kercoh}) we obtain:
\begin{equation}
\int dx d\mu d\nu \quad w(x,\mu,\nu)\langle\alpha|\hat K_{\mu,\nu}|\beta\rangle=
e^{-\frac{|\alpha|^2}{2}-\frac{|\beta|^2}{2}},
\end{equation}
which exactly is the product $\langle\alpha|0\rangle\langle 0|\beta\rangle$ as we expected. As
another example we consider the Schr\"odinger cat state \cite{dod74} of the type discussed in
Ref. \cite{vogel}
\begin{equation}\label{cat}
|\Psi\rangle=\frac{|a+ib\rangle+|a-ib\rangle}{\{2[1+\cos(2ab)\exp(-2b^2)]\}^{1/2}},
\end{equation} 
with $a$ and $b$ arbitrary real numbers,
for which we have a marginal distribution given by \cite{MMT}
\begin{eqnarray}\label{corr2}
w(x,\mu,\nu)&=&\Bigl(\frac{2}{\pi}\Bigr)^{1/2}\Bigl[\frac{1}{\mu^2+\nu^2}\Bigr]^{1/2}
\frac{1}{[1+\cos(2ab)\exp(-2b^2)]}\exp\Bigl[-2\frac{(x-\mu a)^2+b^2
\nu^2}{\mu^2+\nu^2}\Bigr]\nonumber\\ &\times&\Bigl\{\cosh\Bigl[\frac{4\nu b(x-\mu
a)}{\mu^2+\nu^2}\Bigr]+ \cos\Bigl[\frac{2b(2\mu
x-a(\mu^2-\nu^2))}{\mu^2+\nu^2}\Bigr]\Bigr\}.
\end{eqnarray}
Then, from Eq. (\ref{density}) we have in the coherent states basis of Eq. (\ref{kercoh}) 
\begin{eqnarray}
&\int& dx d\mu d\nu\quad w(x,\mu,\nu)\langle\alpha|\hat
K_{\mu,\nu}|\beta\rangle\nonumber\\
&=&\frac{e^{-(a^2+b^2)-(|\alpha|^2+|\beta|^2)/2}}{[1+\cos(2ab)e^{-2b^2}]}
[e^{(\alpha^*+\beta)(a+ib)}+e^{\alpha^*(a+ib)+\beta(a-ib)}+e^{\alpha^*(a-ib)+\beta(a+ib)}+
e^{(\alpha^*+\beta)(a-ib)}]
\end{eqnarray}
which exactly is the product $\langle\alpha|\Psi\rangle\langle\Psi|\beta\rangle$ as we expect.
Finally, we consider the application of Eq. (\ref{density}) to the thermal equilibrium
state of the oscillator at temperature $T$. In this case the Wigner function is given by \cite{gar}:
\begin{equation}
W_T(q,p)=2\tanh\big(\frac{\hbar\omega}{2kT}\big)\exp\big[-(q^2+p^2)
\tanh\big(\frac{\hbar\omega}{2kT}\big)\big].
\end{equation}
From this equation the marginal distribution can be easily obtained  as follows \cite{MMT}
\begin{eqnarray}\label{marther}
w(x,\mu,\nu)&=&\frac{1}{2\pi}\frac{1}{\mu}\int dp\quad 
W_T(\frac{x}{\mu}-\frac{\nu}{\mu}p,p)\nonumber\\
&=&\big[\frac{\Lambda}{\pi(\mu^2+\nu^2)}\big]^{1/2}
\exp\bigl[-\Lambda\frac{x^2}{\mu^2+\nu^2}\bigr],
\end{eqnarray}
where we have introduced $\Lambda=\tanh\big(\frac{\hbar\omega}{2kT}\big)$.
Now, we use Eqs. (\ref{kercoh}) and (\ref{marther}) to calculate 
\begin{eqnarray}
&\int& dx d\mu d\nu\quad w(x,\mu,\nu)\langle\alpha|\hat
K_{\mu,\nu}|\beta\rangle\nonumber\\
&=&2\bigl[\frac{\Lambda}{1+\Lambda}\bigr]\exp\bigl[
\frac{1-\Lambda}{1+\Lambda}\alpha^*\beta-\frac{|\alpha|^2}{2}-\frac{|\beta|^2}{2}\bigr].
\end{eqnarray}
This result is the same that we could have by directly performing 
$\langle\alpha|\hat\rho_T|\beta\rangle$ with the thermal density operator $\hat\rho_T$ given by 
\cite{gar}
\begin{equation}
\hat\rho_T=(1-e^{-\hbar\omega/kT})\sum_n|n\rangle\langle n|e^{-n\hbar\omega/kT}.
\end{equation}

\section{Comparision with the usual tomographic technique}

A relation betweeen the density operator and the marginal distribution analogous to that of Eq.
(\ref{density}) can be derived starting from another operator identity such as \cite{cahgla69}
\begin{equation}\label{rho2}
\hat\rho=\int \frac{d^2\alpha}{\pi}\hbox{Tr}\{\hat\rho\hat D (\alpha)\}D^{-1}(\alpha)
\end{equation}
which, by the change of variables $\mu=-\sqrt{2}\Im m\,\alpha,\quad\nu=\sqrt{2}\Re e\,\alpha$,
becomes
\begin{equation}\label{rho3}
\hat\rho=\frac{1}{2\pi}\int d\mu d\nu\quad \hbox{Tr}\{\hat\rho e^{-i\hat X}\}e^{i\hat X}
=\frac{1}{2\pi}\int d\mu d\nu\quad \hbox{Tr}\{\hat\rho e^{-i\hat x}\}e^{i\hat x}
\end{equation}
where $\hat x=\hat X-\delta$. The trace can be now evaluated
using the complete set of eigenvectors $\{|x\rangle\}$ of the operator
$\hat x$, obtaining
\begin{equation}\label{trace}
\hbox{Tr}\{\hat\rho e^{-i\hat x}\}=\int dx\quad w(x,\mu,\nu)e^{-ix}
\end{equation}
then, putting this one into Eq. (\ref{rho3}), we have a relation of the same form of Eq.
(\ref{density}) with the kernel given by
\begin{equation}\label{kerbis}
\hat K_{\mu,\nu}=\frac{1}{2\pi}e^{-ix}e^{i\hat x}=
\frac{1}{2\pi}e^{-ix}e^{i\mu{\hat q}+i\nu{\hat p}},
\end{equation}
which is the same of Eq. (\ref{ker}) setting $z=1$. It means that we now have 
only one particular Fourier component due to the particular change of variables (the most general
should be $z\mu=-\sqrt{2}\Im m\,\alpha$ and $z\nu=\sqrt{2}\Re e\,\alpha$).

In order to reconstruct the usual tomographic formula for the homodyne detection \cite{APL} we
need to pass in polar variables, i.e. $\mu=-r\cos\phi,\quad\nu=-r\sin\phi$,
then 
\begin{equation}
\hat x\to -r\hat x_{\phi}=-r[\hat q\cos\phi+\hat p\sin\phi].
\end{equation}
Furthermore, denoting with $x_{\phi}$ the eigenvalues of the quadrature operator $\hat x_{\phi}$, we
have
\begin{equation}
\hbox{Tr}\{\hat\rho e^{-i\hat x}\}=\hbox{Tr}\{\hat\rho e^{ir\hat x_{\phi}}\}=
\int dx_{\phi}\quad w(x_{\phi},\phi)e^{irx_{\phi}}
\end{equation}
and thus, from Eq. (\ref{rho3})
\begin{equation}
\hat\rho=\int d\phi dx_{\phi}\quad w(x_{\phi},\phi)\hat K_{\phi}
\end{equation}
with
\begin{equation}\label{kerAPL}
\hat K_{\phi}=\frac{1}{2\pi}\int dr\quad re^{ir(x_{\phi}-\hat x_{\phi})}
\end{equation}
which is the same of Ref. \cite{APL}.
Substantially, the kernel of Eq. (\ref{kerAPL}) is given by the radial integral of the kernel of Eq.
(\ref{kerbis}), and this is due to the fact that we go from a general transformation, with two free
parameters, to a particular transformation (homodyne rotation) with only one free parameter and then
we need to integrate over the other one.

\section{Two-Mode Tomography}

In this Section we discuss an extension of the above approach to multimode systems and, for
simplicity and applications, we concentrate on the two-mode case. From now on we use the
convention that the vector symbol represents only two dimensional vectors.

\subsection{Two-Mode Quasiprobability and Marginal Distribution for Two Squeezed and Shifted
Quadratures}

We introduce the vector operators
$\hat{\vec
q}=(\hat q_1,\hat q_2),\,\hat{\vec p}=(\hat p_1,\hat p_2)$ and then the following
observables  
\begin{equation}\label{5.1}
(\hat{\vec X},\hat{\vec Y})
=\Lambda\;(\hat{\vec q},\hat{\vec p})^T+
(\vec\delta,\vec\delta')
\end{equation}
in which $\Lambda$ is a real symplectic $4\times 4$ matrix
\begin{equation}\label{5.2}
\Lambda\sigma\Lambda^T=\sigma;\quad\sigma =
\left (\begin{array}{clcr} 0&0&1&0
\\0&0&0&1
\\-1&0&0&0
\\ 0&-1&0&0\end{array}\right )
\end{equation}
and $(\vec\delta,\vec\delta')=(\delta_1,\delta_2,\delta_1',\delta_2')$ is
a real c-number four-vector corresponding to the shifts of the four quadratures. Furthermore, the
components
$X_1,X_2,Y_1,Y_2$ are related to the inhomogeneous
symplectic group
$ISp(4,R)$. 
Then we introduce the marginal distribution for the observable $\hat{\vec X}=(\hat X_1,\hat X_2)$
due to relations which, obviously, are two-mode generalizations of the connection between the
characteristic function and the marginal distribution, given in Ref. \cite{cahgla69} for one-mode
case. We have firstly the characteristic function as
\begin{equation}\label{caractfunc}
\chi({\vec k})=\langle e^{i{\vec k}\hat{\vec X}}\rangle=\hbox{Tr}\{\hat\rho e^{i{\vec k}\hat{\vec
X}}\};
\quad {\vec k}\hat{\vec X}=k_1\hat X_1+k_2\hat X_2
\end{equation}
where ${\vec k}=(k_1,k_2)$, and $\hat\rho$ is the density operator of the two-mode system. In terms
of the Wigner function $W({\vec q},{\vec p})$ of the system it is
\begin{equation}\label{caracWig}
\chi({\vec k})=\frac{1}{(2\pi)^2}\int e^{i{\vec k\vec X}}W({\vec q},{\vec p})\, d{\vec q}d{\vec p}.
\end{equation}
Here $\hat{\vec X}$ is, in principle, any observable, but to be concrete below we will consider
this vector to have components
\begin{eqnarray}\label{Xcomp}
&\hat X_1&=\vec\mu\hat{\vec q}+\vec\nu\hat{\vec p}+\delta_1\nonumber\\
&\hat X_2&=\vec\mu'\hat{\vec q}+\vec\nu'\hat{\vec p}+\delta_2
\end{eqnarray}
with $\vec\mu=(\Lambda_{11},\Lambda_{12});\,\vec\nu=(\Lambda_{13},\Lambda_{14});
\,\vec\mu'=(\Lambda_{21},\Lambda_{22});\,\vec\nu'=(\Lambda_{23},\Lambda_{24})$. 
The marginal distribution $w(\vec X,\vec\mu,
\vec\nu,\vec\mu',\vec\nu',\vec\delta)$ of the vector variable $\vec X$, depending on the parameters
of the symplectic transformation $\Lambda$ and shift vector $\vec\delta$, 
is given by the relation
\begin{eqnarray}\label{margX1X2}
w(\vec X,\vec\mu,\vec\nu,\vec\mu',\vec\nu',
\vec\delta)&=&\frac{1}{(2\pi)^2}\int d{\vec k}
\chi({\vec k})e^{-i{\vec k\vec X}}\nonumber\\
&=&\frac{1}{(2\pi)^4}\int d{\vec k}d{\vec q}d{\vec p}W({\vec q},{\vec p})
\exp [-ik_1(X_1-\vec\mu\vec q-\vec\nu\vec p-\delta_1)\nonumber\\
&-&ik_2(X_2-\vec\mu'\vec q-\vec\nu'\vec p-\delta_2)].
\end{eqnarray}
From this formula it is clear that also in the two-mode case the marginal distribution depends on the
difference $\vec X-\vec\delta=\vec x$, so that the replacement $w(\vec
X,\vec\mu,\vec\nu,\vec\mu',\vec\nu',\vec\delta)\to w(\vec x,\vec\mu,\vec\nu,\vec\mu',\vec\nu')$ is
possible, with the normalization
\begin{equation}\label{norm}
\int w(\vec x,\vec\mu,\vec\nu,\vec\mu',\vec\nu')\, d{\vec x}=1.
\end{equation}
Our aim is to express the Wigner function $W({\vec q},{\vec p})$ in terms of the marginal
distribution probability $w(\vec x,\vec\mu,\vec\nu,\vec\mu',\vec\nu')$, i.e. to
invert the formula (\ref{margX1X2}). For this purpouse we perform a Fourier transforms of Eq.
(\ref{margX1X2}) and, after some minor algebra \cite{MMT}, we arrive
at 
\begin{equation}\label{wig1}
W(-{\vec m}/z_1,-{\vec n}/z_1)=(2\pi)^4z_1^4
e^{-i(\vec\mu'\vec m+\vec\nu'\vec n)z_2/z_1} 
w(\vec z,\vec m,\vec n,\vec\mu',\vec\nu') 
\end{equation}
where the vectors $\vec m=(m_1,m_2);\,\vec n=(n_1,n_2);\,\vec z=(z_1,z_2)$ are the
conjugate variables to $\vec\mu$, $\vec\nu$ and $\vec x$ respectively.

\subsection{Two-Mode Quasiprobability and Marginal Distribution for One Squeezed and Shifted
Quadrature}

We will now discuss the connection between the Wigner function $W({\vec q,\vec p})$ and the marginal
distribution $\tilde w(x_1,\vec\mu,\vec\nu)$ for only one quadrature $\hat X_1$. This
latter is related to the previous by
\begin{equation}\label{relmar}
\tilde w(x_1,\vec\mu,\vec\nu)=\int
w(x_1,x_2,\vec\mu,\vec\nu,\vec\mu',\vec\nu')\,dx_2.
\end{equation}
Inserting in this one the expression of $w$ given in Eq. (\ref{margX1X2}), we may see that the 
marginal distribution $\tilde w$ is connected with the characteristic function
$\tilde\chi(k_1)=\chi(k_1,k_2=0)$; where $\chi(k_1,k_2)$ is given in Eq. (\ref{caractfunc}).
Repeating step by step the procedure that leads to Eq. (\ref{wig1}) we have in this case
\begin{equation}\label{wig2}
W(-\vec m/z_1,-\vec n/z_1)=(2\pi)^3z_1^4
\tilde w(z_1,\vec m,\vec n). 
\end{equation}
This formula is similar to the one-mode case discussed in Ref. \cite{MMT} and in the previous 
section,
Eq. (\ref{wig}). The possibility to relate the Wigner function of the two-mode system with different
marginal distributions, $w(\vec x)$ or $\tilde w(x_1)$, is connected with the use of different
quantities of information obtainable from measurements. Of course, the two-dimensional distribution
function contains much more experimental information than the one-dimensional. However, in both
cases, information are overcomplete to determine the Wigner function of the quantum state, but must 
be selfconsistent. In concrete situations these varieties of descriptions might give a freedom to
choose the convenient type of measurements.

\section{Invariant Expression for Two-Mode Density Operator}

We start from a generalization of Eq. (\ref{rho1}) to the two-mode case
\begin{equation}\label{rhobimode}
\hat{\rho}=\int \frac{d^2\alpha_1d^2\alpha_2}{\pi^2} W(\vec\alpha)\hat
T(\vec\alpha)
\end{equation}
with $\vec\alpha=(\alpha_1,\alpha_2)$ and $\alpha_j=(q_j+ip_j)/\sqrt{2};\,j=1,2$.
Then inserting in this one the Wigner function of Eq. (\ref{wig1}), and expressing the
marginal distribution in terms of its Fourier transform, we have
\begin{equation}\label{rhobimodeexp}
\hat{\rho}=\int\frac{d\vec q d\vec p}{(2\pi)^4}\hat T({\vec q},{\vec p})\int 
dx d\vec\mu d\vec\nu \,
z_1^4e^{iz_2(\vec\mu'\vec q+\vec\nu'\vec p)}
w(\vec x,\vec\mu,\vec\nu,\vec\mu',\vec\nu')
e^{iz_1\vec\mu\vec q+iz_1\vec\nu\vec p-i\vec z\vec x}
\end{equation}
that can be written as 
\begin{equation}\label{rhomarg}
\hat{\rho}=\int 
d\vec x d\vec\mu d\vec\nu \,
w(\vec x,\vec\mu,\vec\nu,\vec\mu',\vec\nu')
\hat K_{\vec\mu,\vec\nu,\vec\mu',\vec\nu'}
\end{equation}
with the kernel operator explicitely given by
\begin{eqnarray}\label{K1}
&\hat{K}&_{\vec\mu,\vec\nu,\vec\mu',\vec\nu'}
=\frac{1}{(2\pi)^2}z_1^4e^{-i\vec z\vec x}\nonumber\\
&\times&\exp\Bigl\{\frac{1}{\sqrt{2}}[z_1(
\vec\nu+i\vec\mu)+z_2(\vec\nu'+i\vec\mu')]\hat{\vec a}- 
\frac{1}{\sqrt{2}}[z_1(\vec\nu-i\vec\mu)+z_2(\vec\nu'-i\vec\mu')]
\hat{\vec a}^{\dag}\Bigr\}
\end{eqnarray}
with $\hat{\vec a}=(\hat a_1,\hat a_2)$ the vector operator describing the two modes. In
the case where the Wigner function is given by the marginal distribution of only one
quadrature, as in Eq. (\ref{wig2}), the same procedure leads to an expression similar to
Eq. (\ref{rhomarg}), but with a different kernel operator 
\begin{equation}\label{K2}
\hat{K}_{\vec\mu,\vec\nu}
=\frac{1}{(2\pi)^2}z_1^4e^{-i
z_1x_1}
\exp\Bigl\{\frac{z_1}{\sqrt{2}}[(\vec\nu+i\vec\mu)\hat{\vec
a}-(\vec\nu-i\vec\mu)\hat{\vec a}^{\dag}]\Bigr\}
\end{equation}
which is very similar to that of Eq. (\ref{ker}). It should be noticed that as a direct extension of
the arguments of Sec. 2, also the kernels in Eqs. (\ref{K1}) and (\ref{K2}) are not bounded in the
quadrature (or coordinate) representation, then in this case it is not possible to sample the
density matrix elements.

\section{Applications}

Here we shall consider some applications of the  presented tomographic scheme. We are aware that
the crucial point might be the practical achievement of the generic linear quadrature such as in
Eq. (\ref{Xcomp}). To consider, however, a measurement scheme as an optical implementation of the
developped formalism, let us go back, at first, to the one dimensional case. The quadrature of Eq.
(\ref{quadef}) could be experimentally accessible by using for example the squeezing
pre-amplification (pre-attenuation) of a field mode which is going to be measured (a similar method
in different context was discussed in Ref. \cite{leopauPRL}). In fact, let
$\hat a$ be the signal field mode to be detected, when it passes through a squeezer it becomes
${\hat a}_s={\hat a}\cosh s-{\hat a}^{\dag}e^{i\theta}\sinh s$, where $s$ and $\theta$ characterize
the complex squeezing parameter
$\zeta=se^{i\theta}$
\cite{LK}. Then, if we subsequently detect the field by using the balanced homodyne scheme, we get
an output signal proportiopnal to the average of the following quadrature
\begin{equation}\label{N1-N2}
{\hat E}(\phi)=\frac{1}{\sqrt{2}}({\hat a}_se^{-i\phi}+{\hat
a}^{\dag}_se^{i\phi})\,,
\end{equation}
where $\phi$ is the local oscillator phase. When this phase is locked to
that of the squeezer, such that $\phi=\theta/2$, Eq. (\ref{N1-N2}) becomes
\begin{equation}\label{quamea}
{\hat E}(\phi)=\frac{1}{\sqrt{2}}\left({\hat a}e^{-i\theta/2}[\cosh s-\sinh s]
+{\hat a}^{\dag}e^{i\theta/2}[\cosh s-\sinh s]\right)\,,
\end{equation}
which, essentially, coincides with Eq. (\ref{quadef}), if one recognizes the independent parameters
\begin{equation}\label{munumeas}
\mu=[\cosh s-\sinh s]\cos(\theta/2);\quad\nu=[\cosh s-\sinh s]\sin(\theta/2)\,.
\end{equation}
The shift parameter $\delta$ has not a real physical meaning, since it causes only a displacement of
the distribution along the $X$ line without changing its shape, as can be evicted from Eqs.
(\ref{margXdelta}) and (\ref{margx}). So, in a practical situation it can be omitted.

In the two-mode case 
it could be interesting to use the connection between the Wigner function
and the marginal distribution of only one quadrature in the case of heterodyne detection
(particulary used to detect multimode squeezed states). We may refer to this scheme as the optical
heterodyne tomography. In fact, in balanced heterodyne detection the measured current is determined by
\cite{LK}
\begin{equation}\label{heter}
\hat E(\phi)=\frac{E_1}{\sqrt{2}}[e^{i\phi}\hat a_1^{\dag}+e^{-i\phi}\hat a_1]+
\frac{E_2}{\sqrt{2}}[e^{i\phi}\hat a_2^{\dag}+e^{-i\phi}\hat a_2]\,.
\end{equation} 

If one uses squeezed pre-amplification
(pre-attenuation) for each single mode, as discussed above,
both $E_1$ and $E_2$ are indipendent variables,
however, we now have only one
phase, the local oscillator phase $\phi$; thus, in order to have a fourth independent
variable which would allow a complete reconstruction  of two-mode quasi-probability, we could
consider the phase shifter used in the first step of the measurement procedure. Since the two modes
have different frequencies, the phase shifter induces different phase changes in the two modes; for
example
\begin{equation}\label{phaseshift}
{\hat a}_1\to{\hat a}_1e^{i\theta_1}\,;\quad{\hat a}_2\to{\hat a}_2e^{i\theta_2}\,;
\end{equation}
where the difference $\theta_1-\theta_2$ depends on the length of the optical path in the phase
shifter. Upon using this requirement, the measured quadrature (\ref{heter}) will become
\begin{equation}\label{hetquad}
\hat E(\phi)=\frac{E_1}{\sqrt{2}}[e^{i(\phi+\theta_1)}\hat a_1^{\dag}+e^{-i(\phi+\theta_1)}\hat
a_1]+
\frac{E_2}{\sqrt{2}}[e^{i(\phi+\theta_2)}\hat a_2^{\dag}+e^{-i(\phi+\theta_2)}\hat a_2]\,,
\end{equation} 
that corresponds to the quadrature $X_1$ of Eq. (\ref{Xcomp}) with 
$\vec\mu=(E_1\cos(\phi+\theta_1),E_2\cos(\phi+\theta_2))$ and
$\vec\nu=(E_1\sin(\phi+\theta_1),E_2\sin(\phi+\theta_2))$.
For the shift parameter $\delta_1$ the previous considerations, made in one mode case, still
hold. 

We would now present some examples of the probability $\tilde w$ measurable with the
above scheme. We first consider 
the most general two-mode squeezed states described by the Wigner funtion of the
form \cite{DMM}
\begin{equation}\label{Wsqueez} 
W({\vec q,\vec p})=(\det{\cal
M})^{-1/2}\exp\Bigl[-\frac{1}{2}\Bigl(({\vec q,\vec p})-\langle (\hat{\vec q},\hat{\vec
p})\rangle\Bigr) {\cal M}^{-1}\Bigl(({\vec
q,\vec p})-\langle (\hat{\vec q},\hat{\vec p})\rangle\Bigr)^T\Bigr]
\end{equation} where the real symmetric dispersion matrix $\cal M$ has ten variances
\begin{equation}\label{var} 
{\cal M}_{\alpha\beta}=\frac{1}{2}\langle (\hat{\vec q},\hat{\vec p})_{\alpha}(\hat{\vec
q},\hat{\vec p})_{\beta}+
(\hat{\vec q},\hat{\vec p})_{\beta}(\hat{\vec q},\hat{\vec p})_{\alpha}\rangle - \langle
(\hat{\vec q},\hat{\vec p})_{\alpha}\rangle\langle (\hat{\vec q},\hat{\vec p})_{\beta}\rangle
\qquad\alpha,\beta=1,2,3,4.
\end{equation} 
By integrating only on $k_1$ with $k_2=0$, we obtain the
marginal distribution of only one quadrature $X_1$ as given in Eq. (\ref{margX1X2})
\begin{eqnarray}\label{margsqueez} 
{\tilde w}(x_1,\vec\mu,\vec\nu)&=&\frac{1}{(2\pi)^3}\int
dk_1\int d{\vec q} d{\vec p}W({\vec q,\vec p}) e^{-ik_1(x_1-\vec\mu{\vec q}
-\vec\nu{\vec p})}\nonumber\\
&=&[2\pi(\vec\mu,\vec\nu){\cal M}
(\vec\mu,\vec\nu)^T]^{-1/2}\exp\Biggl[-\frac{x_1^2} {2
(\vec\mu,\vec\nu){\cal M}
(\vec\mu,\vec\nu)^T}\Biggr]
\end{eqnarray} 
where we have assumed 
$\langle (\hat{\vec q},\hat{\vec p})_{\alpha}\rangle=0$ for all $\alpha$. 
This distribution is a Gaussian for the quadrature variable.

As a second example, we consider the two-mode Schr\"odinger-cat state
\cite{Ansari} 
defined as
\begin{equation}\label{cats}
|{\vec A}_{\pm}\rangle=N_{\pm}(|{\vec A}\rangle\pm |-{\vec A}\rangle)
\end{equation}
with 
\begin{equation}\label{catsnorm}
N_+=\frac{e^{|{\vec A}|^2/2}}{2\sqrt{\cosh |{\vec A}|^2}};\;
N_-=\frac{e^{|{\vec A}|^2/2}}{2\sqrt{\sinh |{\vec A}|^2}}.
\end{equation}
The Wigner function of this state is
\begin{eqnarray}\label{catwig}
W_{{\vec A}_{\pm}}({\vec q,\vec p})=N_{\pm}^2&[&W_{(\vec A,\vec B=\vec A)}({\vec q,\vec p})\pm
W_{(\vec A,\vec B=-\vec A)}({\vec q,\vec p})
\nonumber\\
&\pm &W_{(-\vec A,\vec B=\vec A)}({\vec q,\vec p})\pm W_{(-\vec A,\vec B=-\vec A)}({\vec q,\vec
p})]
\end{eqnarray}
where
\begin{equation}\label{cohwig}
W_{\vec A,\vec B}=4\exp \Bigl[-2\vec\alpha \vec\alpha^*+2{\vec A}\vec
\alpha^* +2{\vec B^*}\vec\alpha-{\vec A \vec B^*}-\frac{|{\vec A}|^2}{2}-\frac{|{\vec
B}|^2}{2}\Bigr]
\end{equation}
is the Wigner function for two-mode coherent state \cite{ref1lectures} with 
\begin{equation}
\vec\alpha=\frac{{\vec q}+i{\vec p}}{\sqrt{2}}.
\end{equation}
Let us now consider the even coherent state, then to get the marginal distribution, we use, as in
the previous case, Eq. (\ref{margX1X2}) obtaining 
\begin{eqnarray}\label{intgauss2}
&\tilde{w}&(x_1,\vec\mu,\vec\nu)=
\frac{2\pi^{-1/2}N_+^2}{\sqrt{|\vec\mu|^2+|\vec\nu|^2}}
\exp\Bigl[\frac{-x_1^2-(\nu_1P_1+\mu_1Q_1)^2-(\nu_2P_2+\mu_2Q_2)^2}
{|\vec\mu|^2+|\vec\nu|^2}\Bigr]\nonumber\\
&\times&\Bigl\{\exp\Bigl[
\frac{-(P_1^2+Q_1^2)(\nu_2^2+\mu_2^2)-(P_2^2+Q_2^2)(\nu_1^2+\mu_1^2)+
2(\mu_1P_1-\nu_1Q_1)(\mu_2P_2-\nu_2Q_2)}
{|\vec\mu|^2+|\vec\nu|^2}\Bigr]\nonumber\\
&\times&\cos\Bigl[\frac{2(\mu_1P_1+\mu_2P_2-\nu_1Q_1-\nu_2Q_2)x_1}
{|\vec\mu|^2+|\vec\nu|^2}\Bigr]\nonumber\\
&+&\exp\Bigl[\frac{-2(\nu_1P_1+\mu_1Q_1)(\nu_2P_2+\mu_2Q_2)}
{|\vec\mu|^2+|\vec\nu|^2}\Bigr]
\cosh\Bigl[\frac{2(\nu_1P_1+\nu_2P_2+\mu_1Q_1+\mu_2Q_2)x_1}
{|\vec\mu|^2+|\vec\nu|^2}\Bigr]\Bigr\}.
\end{eqnarray}
where we have set
\begin{equation} {\vec A}=\frac{{\vec Q}+i{\vec P}}{\sqrt{2}}.
\end{equation}
In the case of $\mu_2=\nu_2=0$ and $Q_2=P_2=0$, the marginal distribution of Eq.
(\ref{intgauss2}) reduces to the one-dimensional marginal distribution of Eq. (32) of Ref.
\cite{MMT} \footnote{Up to a misprint in this reference.}.

\section{Conclusions}

We have shown that the optical tomography may be extended to two-mode case,
and the new observables introduced, which are generic linear forms in quadratures, give the
possibility of using various measurements, different from the usual ones, to determine
the Wigner function of the system under study in terms of the marginal distributions.
To better understand the meaning of the measure of the observable $\hat X$ of Eq.
(\ref{quadef}), we recall that this symplectic transformation could be represented as a
composition of shift, rotation and squeezing. 
Along this line we have presented some schemes which could be implemented in optics.
To be more precise, the shift parameter does not play a real physical role in the measurement process,
it has been introduced for formal completeness and it expresses the possibility to achieve the desired
marginal distribution by performing the measurements in an ensemble of frames which are  each
other shifted; (related method was early discussed in Ref.
\cite{Royer}). In an electro-optical system this only means to have the freedom of using different
photocurrent scales in which the zero is shifted by a known amount.

The tomographic formalism is presented here in an invariant form for the density operator and it
may be suitable for further numerical analysis. It is worth to remark that all the formulae  of
two-mode case may be obviously rewritten for N-mode case with arbitrary N. The presented extension
gives the possibility to include the procedure of heterodyne detection (two-mode case) as well.
Furthermore, the extension of the tomographic technique to multimode systems could also be useful to
obtain the (complete) quantum information about a system in an indirect way \cite{MT}.

Another conclusive remark is related to the group structure of the presented tomographic extension.
Using generic observables which are linear in quadratures, we apply a symplectic group
transform to an initial quadrature component. In one mode optical tomography this transform belongs
to the rotation subgroup of the symplectic group $ISp(2,R)$. Thus the presented construction of the
Wigner function (density operator) may be reformulated as the group problem for a particular
symplectic group and we could call our extension as "symplectic optical tomography". On the other
hand the scheme might be generalized to other Lie groups different from the symplectic one. We
will develop these points in future papers.

\section*{Acknowledgments}
This work has been partially supported by European Community under the Human Capital and Mobility
(HCM) programme. Part of the work was developped at Corvara during the annual meeting of the HCM
Network "Non-classical Light". V.I.M deeply acknowledges the financial support of Istituto Nazionale
di Fisica Nucleare. P. T. acknowledges M. G. Raymer to bring his attention on Ref. \cite{Royer}.

\end{document}